\documentclass[preprint2]{aastex}		

   \newcommand{\nd}{\nodata}
   \newcommand{\tnm}{\tablenotemark}
   \newcommand{\gsim}{\rlap{$>$}{\lower 1.0ex\hbox{$\sim$}}}

   \shorttitle{HST Morphologies of K-selected EROs}
   \shortauthors{Yan \&\ Thompson}

   \normalsize

\begin{document}

\title{HST/WFPC2 Morphologies of K-selected \\
       Extremely Red Galaxies}

\author{Lin Yan}
\affil{SIRTF Science Center, MS~220-6, Caltech, Pasadena, CA~91125}
\email{lyan@ipac.caltech.edu}

\author{David Thompson}
\affil{The Palomar Observatories, Caltech, Pasadena,  CA~91125}
\email{djt@irastro.caltech.edu}

\begin {abstract}

We selected 115 extremely red objects (EROs) from deep HST WFPC2 archive 
data combined with ground-based K-band images, with $\rm (F814W - K_s) 
\geq 4$, K-band $\rm SNR \geq 5$, and a median limiting K$_s$ magnitude 
of $\sim 18.7$, over a corresponding area of 228 sq. arcminutes, for a 
morphological study of the ERO galaxy population.  The survey covered a 
total of $\sim$409 sq. arcminutes over 77 separate WFPC2 fields.  
This is the first complete sample of bright EROs with high resolution 
HST morphologies.  From a visual morphological classification, we find 
that 30$\pm$5\% of our $\rm (F814W - K_s) \geq 4$ selected sample 
have morphologies consistent with a pure bulge or 
bulge-dominated galaxy (equivalent to E/S0), while disks comprise 
64$\pm$7\% of the sample.  Only 6\%\ of the EROs remained unclassifiable.  
Mergers or strongly interacting systems, which includes sources from 
both classes, make up 17$\pm$4\% of the full sample.  The quantitative 
MDS profile fitting is consistent with these results.  These results 
highlight the complex nature of optical/near-IR color selected EROs.  
The dominant component of our sample is comprised of disks, not spheroids 
or strongly interacting systems like HR\,10.  Using Bruzual \& 
Charlot SED models, we investigated population differences in EROs 
selected by their ($I-K$) vs. ($R-K$) colors and found that I-band 
based surveys preferentially select systems with prolonged star formation.  
Real differences in the surface densities of EROs in $R$-band and $I$-band 
based survey may reflect this color selection effect, complicating the 
comparisons between and interpretations on the nature of the ERO population.  
We conclude that only a small fraction of EROs at $z\sim 1 - 2$ could 
be passively evolving ellipticals formed at high redshift through a 
``monolithic collapse'' mechanism.  For the majority of EROs, even 
if most of their stellar mass is already in place at $z \sim 1$, 
interaction with the environment and accretion of gas still play important 
roles in their continuing evolution.

\end {abstract} 

\keywords{galaxies: elliptical -- 
          galaxies: fundamental parameters -- 
          galaxies: high-redshift -- 
          galaxies: spiral}

\section{Introduction}

Extremely Red Objects (EROs), which have optical-to-infrared 
colors which differ significantly from typical field sources, 
encompass a wide variety of phenomen\ae.  Galaxies of assorted types 
make up the dominant component of ERO samples, but one can also find 
low mass stars, gravitationally lensed sources, and transient 
sources such as variable stars, asteroids or supernovae which may 
not be initially recognized as such.  The term Extremely Red 
Galaxies (ERGs) is also commonly used, sometimes interchangeably, but 
usually refers to a sample of EROs which has been cleaned of the objects 
which are not galaxies.  ERGs are therefore a subset of the EROs.  We 
adopt the more general term ERO throughout this paper.

Both the definition and interpretation of EROs has evolved somewhat 
since their initial discovery, and it is useful to review the subject 
here for some historical perspective.  When first identified as a 
distinct population of sources ~\citep{ERR88}, EROs were thought to be 
good candidates for primeval galaxies.  Subsequent observations 
~\citep{ERR89}, however, showed these early EROs to be $z \sim 0.8$ 
elliptical galaxies.  Additional EROs were noted in the following years 
\citep{MPW92,ED92,Pea93,Graham94,HR94,Sea94,DSD95,Dj95,Treu98,Im02}.  Most 
of these were serendipitous detections, identified on images targeting known, 
high-redshift radio galaxies or other active galactic nuclei.  Little or 
no followup work was done on these objects at the time, which to some 
extent reflected the limited capabilities of existing instruments and 
telescopes.  These EROs were identified with colors spanning a wide range: 
($R-K$)\,$\geq$\,5--7, or ($I-K$)\,$\geq$\,4--6.  

A resurging interest in EROs accompanied the development of the 
Submillimeter Common User Bolometer Array \citep[SCUBA]{SCUBA}, and 
the subsequent detection of the extremely red galaxy HR\,10 ~\citep[this 
source is also known as HR94\,10 or ERO\,J164502+4626.4]{HR94} 
at 850\,$\mu$m ~\citep{cim98,dey99}.  At a redshift of 1.44 ~\citep{GD96}, 
the detection of HR\,10 in the submillimeter implied the presence of 
massive quantities of dust accompanied by very high star formation rates.  
The ERO population was thought to provide fertile hunting grounds for 
more submillimeter-bright galaxies at high redshift. Additional 
observations have not supported this idea, however, with only a 
relatively small fraction, on the order of 20\% ~\citep[Thompson, 
priv. comm.]{and99,mohan02}, of the bright EROs ($K < 19.5$) showing 
strong submillimeter emission.

The development of larger-format infrared arrays and wider field 
instrumentation enabled subsequent field surveys to cover enough 
area to assemble significant samples of systematically selected EROs 
for further study, in blank fields ~\citep{tho99,daddi00a,LCIR01} as 
well as targeted surveys ~\citep{Chap00,Cim00,Liu00}.  \citet{tho99} 
adopted a color selection for EROs of ($R-K$)\,$>$\,6\fm0.  The 
motivation was that this color was redder than the expected colors of 
elliptical galaxies with anything but the highest formation redshifts 
($z_f > 10$), and thus represents an extreme color for any normal galaxy.  
The assumption at the time was that the extremely red galaxy population 
consisted of either old ellipticals or young, dusty starbursts 
~\citep{cim98,tho99,dey99}.  The relative contribution of these two types 
of galaxies would have a bearing on the timing of massive galaxy formation 
and their subsequent evolution.  It is important to emphasize that, at 
that time, the term ``young, dusty starbursts'' referred specifically to 
massive starbursts like that seen in HR\,10 or luminous infrared galaxies.  
Multi-band photometry could potentially distinguish between ellipticals and 
starbursts ~\citep{PM00}, but this technique requires very low photometric 
uncertainties to work well (see, for example, ~\citet{Mea02}).

In order to better study the $z \sim 1$ elliptical galaxy population, 
\citet{daddi00a} adopted a bluer color selection limit, ($R-K$)\,$>$\,5.3, 
set by the expected colors of a $z = 1$ passively evolving old stellar 
population.  This definition, or the roughly equivalent ($I-K$)\,$>$\,4.0, 
for the ERO color selection criterion has generally been adopted in 
the majority of subsequent work.  

While there are a number of redshifts now known for EROs 
~\citep{GD96,Sea99,Liu00,afonso01,smith02b}, systematic redshift 
surveys of complete samples are only now becoming available 
~\citep{cim02}.  Morphological information based on high-resolution 
HST imaging for complete samples of EROs are also only now 
starting to appear ~\citep[this work]{smith02a}.  Without similar 
spectroscopic or morphological information, earlier ERO surveys divided 
the ERO population into two components: old, evolved systems or dusty, 
massive starbursts.  But the true nature of K-selected EROs is likely 
to be much more complex, as suggested by recent work ~\citep{LCIR01,cim02}. 

~\citet{LCIR01} find a large scatter in the $(V - I)$ colors of their 
ERO sample, best fit by passive evolution models with extended star 
formation ($\tau = 1$~Gyr).  This implies that the star formation 
history of EROs is more complex than a binary division into evolved 
ellipticals or dusty, massive starbursts implies.  From their K20 survey, 
~\citet{cim02} found that about half of their spectroscopic sample of 
$\sim$30 EROs are dusty star-forming galaxies with emission lines, while 
the remaining half are old stellar populations with absorption line 
spectra.  However, the simple presence of line emission could span a 
wide range of galaxy types, from bulge-dominated, late type spiral galaxies 
with a small amount of star formation through the more massive starbursts 
like HR\,10.  Dust could also completely obscure any on-going star 
formation, to the point that the optical and UV emission lines are not 
seen ~\citep{PW00}.  Examples of what appear to be quiescent disks at 
$z \sim 1.5$ exist ~\citep{dokk01,smith02b}.  

There are important differences in the formation and evolution of 
quiescent normal galaxies and young, dusty, massive starbursts.  
Morphologies have the potential to distinguish between the various 
interpretations, which motivated this work.

In this paper, we present the high resolution morphologies
derived from HST WFPC2 images for a large sample of K-selected EROs.
Our results reveal for the first time a new type of ERO which
dominates the population and is neither an early-type galaxy nor a dusty,
massive star forming galaxy.  We will also discuss the implications 
of our results for the past and future evolution of massive galaxies 
at $z \sim 1$. 

\section{Observations and Reductions}

The data used in this survey come from two sources.  Deep, high resolution
HST/WFPC2 F814W images from archival data, specifically from the
Medium Deep Survey (MDS; ~\citet{griff94,rat99}), are used for the optical 
dataset.  Images in the $K_s$ band were obtained from the Palomar 60-inch 
telescope for a total of 77 MDS fields.  

Details on the survey design, observations, reductions and analysis, 
as well as the details of our visual morphological classification and 
the automated two-component profile fitting results from the MDS, are
given below.  In summary, magnitudes were extracted in matched apertures 
after rotating, rebinning, and convolving the HST data to match the 
infrared data.  A final set of 115 unique EROs satisfying the selection 
criteria (F814W - K$_s$) $\geq$ 4 and with a K-band signal-to-noise 
ratio (SNR) $\geq 5$ were identified over a total area of 409 sq. 
arcminutes.  The coordinates, photometry, and morphological classifications 
for the 115 EROs are listed in Table~\ref{ERO_tab}, along with the field name 
and source identification from the Medium Deep Survey.  


\subsection{Field Selection}

Our target fields were selected from the MDS database to have a 
5$\sigma$ sensitivity of F814W$\geq$24$^{\rm m}$ (Vega), which would 
provide a good signal-to-noise ratio (SNR), high resolution optical 
image for morphological classifications for the majority of the EROs 
we might detect.  This corresponds to a minimum total exposure time 
of $\geq 2700$\,s, typically split over 2 or more exposures to aid in 
the removal of cosmic rays.  We restricted the target fields to high 
Galactic latitudes, $b_{II} \geq 20^\circ$, to minimize foreground 
stellar contamination, and make no explicit corrections for Galactic 
extinction.  For most fields, the color correction is under 0\fm1.  
Finally, we selected primarily northern hemisphere fields ($\delta 
\geq -15^\circ$) to preserve accessibility from Palomar Observatory.

The fields were selected without regard to any specific science targets.  
The majority of the MDS database is composed of random parallel fields 
imaged while the primary science target was observed in another 
HST instrument.  However, there are a significant number of fields 
containing known clusters present in the MDS database, twelve of which 
we imaged as part of this survey.  Gravitational lensing from these 
foreground clusters (typically at redshifts of a few tenths) can boost 
the observability of faint EROs.  This technique has been used with 
success in surveys targeting EROs ~\citep{smith02a}, submillimeter-bright 
galaxies ~\citep{smail99}, and other high redshift sources 
~\citep{Ellis01}.  We note here that any lensing will be unbiased with 
respect to galaxy morphologies, and so we make no specific corrections 
for lensing from cluster fields observed in this survey.  Lensing can, 
however, boost the surface densities of EROs (see \S\ref{SDsect}).  We 
note that there is no overlap in target fields between our survey and 
that of ~\citet{smith02a}.

\subsection{WFPC2 F814W Images}

We retrieved the F814W images from the MDS database at Space Telescope 
Science Institute (see \url{http://archive.stsci.edu/mds/cdrom.html}). 
These data have been processed through the MDS data reduction pipeline, 
which includes warm pixel correction, image stacking, removal of cosmic 
rays, and photometric calibration.  For a complete description of the 
MDS reduction pipeline, see ~\citet{rat99}. 
In general, this automated reprocessing produces the best results on 
fields where multiple (n$>$3), dithered images were obtained, and where 
the field and its immediate surroundings are free of bright stars.  
These processed images are stored in the MDS database in the HST standard 
GEIS format, where each of the 4 CCDs and their associated header 
information is stored in a separate layer of the disk file.

For each of our target fields, we retrieved these GEIS format images,
then processed them further to meet the needs of our survey.  The
data were processed using standard IRAF\footnote{IRAF is distributed by
the National Optical Astronomy Observatories, which are operated by
the Association of Universities for Research in  Astronomy, Inc.,
under cooperative agreement with the National Science Foundation.}
tasks, plus specialized routines developed for WFPC data found in the
STSDAS package.

First, we interpolated over bad pixels flagged by the MDS processing, 
subtracted the background, and corrected any additional deviant pixels 
identified by the {\sc cosmicray} routine.  The STSDAS task {\sc wmosaic} 
was used to correct for distortion in the WFPC2 images and assemble the 
data from the four separate CCDs into a single mosaic image.  This additional 
processing does miss some cosmic rays, but any contamination of the EROs 
is minimal, and only serves to make the sources bluer than they would
otherwise be.

\subsection{K-band Imaging} 

Ground-based $K_s$ images were obtained using the near-IR camera on the 
Palomar 60~inch telescope ~\citep{p60ircam}.  The camera has a 
$160^{\prime\prime} \times 160^{\prime\prime}$ field of view, well-matched 
to WFPC2.  The detector is a 256$^2$ pixel HgCdTe, covering 0\farcs62 
per pixel.  A total of 42 nights went into this project, spanning the time 
period of 2000 August 11 -- 2001 June 06 UT.  Useful data was collected on 
21 photometric nights.  We imaged 77 MDS fields under photometric conditions, 
with repeat observations on about one-third of the fields to help determine 
the photometric uncertainties.  The repeat observations show less than 
0\fm1 of systematic variations.  We targeted exposure times of 75 minutes 
(4500 seconds) per field, which yields a 5$\sigma$ point-source sensitivity 
of $K_s \sim$18\fm75 in good seeing.

The data were reduced following standard infrared reduction procedures.
Each frame was sky-subtracted with temporally-adjacent images and
then flatfielded with a combination of dome and twilight sky flats.
Offsets were determined from as many sources as possible (with a minimum
of one, as the header coordinates were insufficiently accurate to
stack the data with confidence).  A bad pixel mask was used to reject
dead or excessively hot pixels, and the data combined with integer
pixel offsets into the final mosaics for each field.

Observations of the Persson infrared photometric standards (Persson 
et al 1996) were obtained throughout the night, and used to derive
the zero points and airmass corrections.  For the photometric data,
the zero points show only 0.02 mag rms variations from night to night.  
This is added into the photometric uncertainties of individual objects 
in quadrature. 

The 5$\sigma$ sensitivities in the centers of the infrared
mosaics ranges from 18\fm2 $-$ 19\fm5.  This was calculated from
the per-pixel sky noise at the center of the infrared mosaics, and
scaled to an aperture diameter equal to 2.5 times the FWHM of the
seeing.  The seeing ranges from 1\farcs2 to 2\farcs5, with the majority
of the data better than 1\farcs8 (3 pixels FWHM).  The lower
sensitivities are generally due to poorer seeing conditions, while the
deeper data are from stacking observations from multiple nights.
A summary of the useful observations are given in Table~\ref{obs_tab},
which lists the field, exposure times for both the $K_s$ and F814W images,
and the central 5$\sigma$ $K_s$ sensitivity.

\subsection{Matched Aperture Photometry}

The original full resolution F814W images were rotated, rebinned 
and convolved to match the infrared image orientation, scale and seeing. 
The optical and infrared images were then co-aligned and cropped to 
matching areas on the sky.  The resulting image pairs were used for 
extraction of the photometry.  A set of IRAF scripts were written to 
speed the execution of these and subsequent tasks in creating the final 
photometric catalogs.

\begin{figure}[!ht]
   \plotone{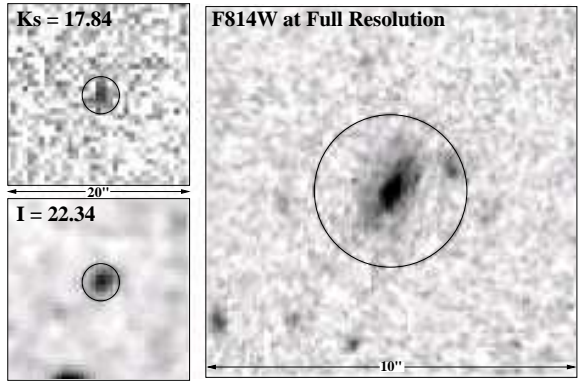}
   \caption{An example from the aperture photometry review process.  The
   photometric images, with the HST data rebinned and convolved to match
   the infrared data, are shown in the two panels to the left.  The larger
   image to the right shows the full-resolution WFPC2 data.  The circles
   in all three panels are scaled to the size of the aperture used for
   photometry (here, 4\farcs2 diameter).  \label{mdsreview}}
\end{figure}

Because the K-band image mosaics have increasing noise at the edges, source 
detection was performed on a noise normalized image so that a uniform 
detection threshold could be applied.  The noise normalization was done 
by multiplying the K-band image mosaic by the square root of an exposure 
time map.  We used the Source Extractor software ~\citep{SExtract}
for the initial source detection on the K-band data.  However, 
uncorrected distortions and/or small residual offsets between the image pairs 
ultimately required recentering on objects in the scaled and seeing-convolved 
HST data.  Re-centering of the photometric apertures on the HST data was 
reviewed interactively for all sources with initial colors greater than 
($F814W-K_s$) $>$ 3\fm8, as illustrated in Figure~\ref{mdsreview}.  This 
review process also allowed for rejection of spurious sources from the 
photometric catalog, as well as noting the effects of uncorrected cosmic 
rays or contamination from nearby galaxies.  Note the complete lack of 
morphological information in the ground-based imaging, while the 
full-resolution image clearly shows a galaxy with a disk.  

The final aperture photometry was extracted using the {\sc apphot} package 
in IRAF.  We adopted an aperture diameter of 2.5 times the seeing FWHM 
of the corresponding K-band image.  This represents a compromise between 
a larger aperture, which yields a fairer measurement of a galaxy's total 
magnitude as well as provides a better measure for intrinsically large 
galaxies, and a smaller aperture, which would minimize contamination from 
other sources close to the line-of-sight to the ERO.  The majority of the 
K-band data are better than 1\farcs8 FWHM (3 pixels), so the photometry was 
extracted in apertures typically smaller than 4\farcs5 in diameter.  
Because the K-band images have non-uniform noise and depth, the 
SNR for each galaxy was determined locally.  For faint objects, 
the SNR is dominated by the sky noise within the photometric 
aperture.  

In 20\%\ of the cases, the photometric apertures do include close companions 
which could affect the F814W photometry.  This has the effect of making 
our measured colors for these objects bluer than the true color of the ERO, 
and thus a lower limit.  We were able to correct for the majority of 
contamination problems during the review process, but we note that the 
presence of companions should not be biased toward any particular 
morphological type.

\subsection{Survey Area}
\label{sec:area}

Because our survey is composed of many separate images obtained on 
different nights, under different seeing conditions, and with different 
exposure times, the depth of each field varies.  Also, since the 
infrared camera field size is only slightly larger than the HST WFPC2 
mosaic, and given the random orientations for the HST data, portions 
of the HST images often overlap the higher-noise area around the edges 
of the infrared mosaics.  These factors combine to make the total 
survey area a function of the depth, while allowing maximal use of the 
survey data.

We integrated the survey area as a function of magnitude on a per 
pixel basis.  Ideally, this should be done over resolution elements.  
However, aside from single-pixel deviations due to masked bad pixels 
or cosmic rays, the depth in the infrared data only changes relatively 
slowly with position around the edges.

To sum the survey area, we first created a mask for the co-aligned images 
covering the area on the sky outside the borders of the WFPC2 data.  An 
effective exposure time map for the infrared mosaic was then converted 
into a map of the limiting magnitude in each pixel as:

\begin{center}
   $m_{pix} = m_{lim,cent} + 2.5{\rm log}(t_{pix}/t_{max})^{0.5}$
\end{center}

\noindent{where $t_{pix}$ is the exposure time that went into the individual 
pixels in the final infrared mosaic, and $t_{max}$ is the total 
exposure time for that image.}

\begin{figure}[!ht]
   \plotone{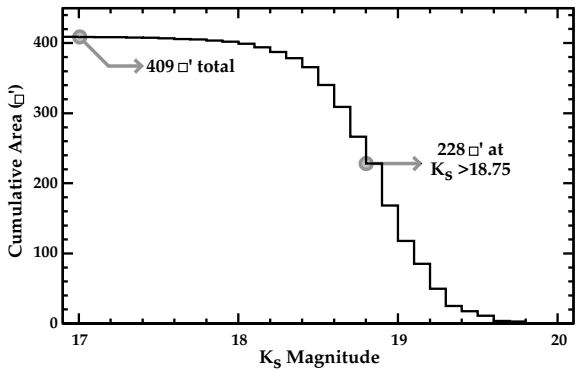}
   \caption{Area covered by this survey as a function of the $K_s$ magnitude.
   Half of the total area was surveyed to at least $K_s=$18\fm75 in depth.
   \label{surveyarea}}
\end{figure}

A histogram of the unmasked pixel limiting magnitudes (pixels also 
covered by the HST data) was then created for each survey image in 
0\fm1 bins.  The resulting 77 histograms were combined, and scaled 
by the area per pixel.  This differential area histogram was finally 
integrated from the faint to the bright limits.  A plot of the integrated 
area as a function of the limiting magnitude is shown in 
Figure~\ref{surveyarea}.  The full survey area is 409 square arcmin, with 
more than half of this (228 sq. arcmin) reaching to at least a depth of 
$K_s=$18\fm75.

\subsection{The ERO Sample}

\begin{figure}[!ht]
\plotone{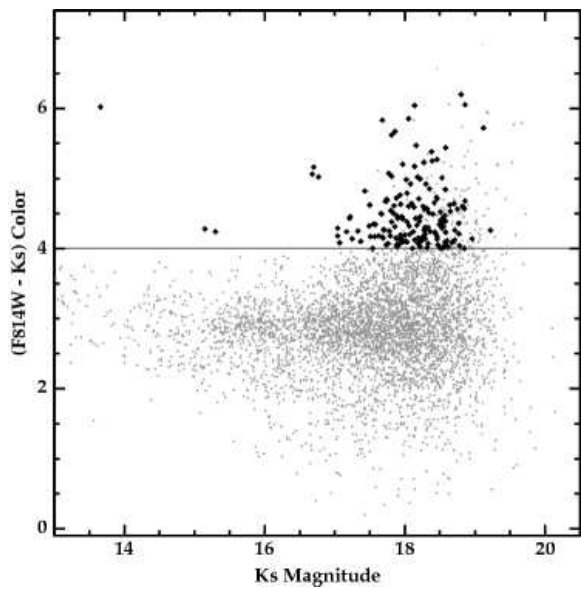}
\caption{The full color-magnitude diagram for this survey.
All points are plotted, including duplicate observations of the same source
on two or more nights.  Sources satisfying the color selection of $\rm
(F814W - K_s) \geq 4$, and with a K-band $\rm SNR \geq 5$, are plotted
with heavier symbols.  \label{fullcmd}}
\end{figure}

\begin{figure}[!ht]
\plotone{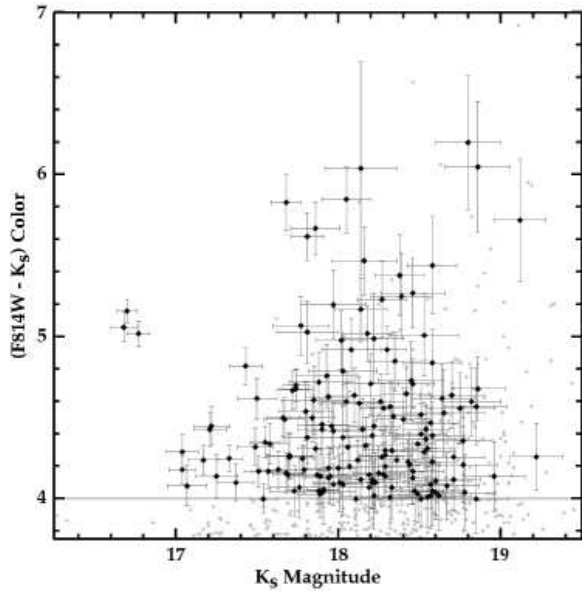}
\caption{A higher resolution plot of the color-magnitude diagram shown
in Figure~\ref{fullcmd}, covering the region where EROs were selected.
Photometric uncertainties for the EROs are also plotted.  Note that there
are a number of sources which are redder than the color selection limit,
but which were not kept in the ERO sample.  Most often, this was due to
a low SNR in the K-band image.  \label{zoomcmd}}
\end{figure}

A total of 115 unique galaxies satisfy the selection criteria of 
(F814W - K$_s$) $\geq$ 4 and with a K-band signal-to-noise ratio 
(SNR) $\geq 5$, were selected over a total area of 409 sq. arcminutes.  
The coordinates, photometry, and morphological classifications for these 
115 EROs are listed in Table~\ref{ERO_tab}, along with the field name 
and source numerical identification from the Medium Deep Survey.  We 
show the combined color-magnitude diagram for all 77 fields in 
Figure~\ref{fullcmd}.  All independent observations are included, so 
some sources are plotted more than once.  In Figure~\ref{zoomcmd} 
we detail the photometric uncertainties on the ERO sample.  

Not included in Table~\ref{ERO_tab} are six sources which were obvious 
stars, most of which were not intrinsically red but simply saturated 
on the HST images.  We also excluded the gravitationally lensed system 
MG 0414$+$0534 ~\citep{MG0414}, which does satisfy the adopted selection 
criteria, but was the science target for the HST observations and thus 
was not a randomly selected source. 

\subsection{Morphological Classification}

The main goal of this survey is to utilize the high resolution WFPC2 
images to study the morphological properties of EROs. We classified 
the sources visually, but also compare our results to the automated 
profile fitting from the MDS.  The details are described below.  

The full-resolution HST/WFPC2 images of all of the EROs in our sample 
are shown in Figure~\ref{hstimgs}.  The authors strongly caution the 
reader against attempting to morphologically classify the EROs solely 
from these grayscale images.  The HST data are freely available from 
STScI, and interested readers are encouraged to retrieve the original 
data.

\subsection{Visual Classification}

Visual morphological classification on nearby galaxies has had a 
significant impact on our understanding of galaxy formation, 
environment, and evolution. However, it is widely acknowledged 
that visual classification is an inherently uncertain and subjective 
process.  In the high redshift regime, the visual classification 
of galaxy morphology is further complicated by limited resolution 
(even with HST), lower SNR, $(1+z)^4$ cosmological dimming, and 
observations at restframe wavelengths which vary with the redshift.  
Several detailed studies on the biases in visual morphological 
classifications of high redshift galaxies have been made 
~\citep{hibvac97,Ode96}. Despite its inherent uncertainties 
and subjectiveness, visual morphological classification has proved 
to be a powerful tool for galaxies at $z \sim 1$, as demonstrated by
by many studies using deep HST images in the HDF, as well as 
morphological studies of galaxies in high-redshift clusters 
~\citep{Lubin98,Dressler97,couch98,smail98}. 
One important point illustrated in these empirical studies, and also 
in simulations \citep{hibvac97}, is that the extended morphological
features remain readily visible in deep HST F814W images out to 
$z \sim 1$.  In light of this, we have visually classified all the 
EROs in our sample.

Because of the bright K$_s$ magnitude limits in this survey, all 115 
galaxies in our sample are detected in the F814W images, most well
resolved and with a high SNR.  As a training set for our visual 
classification, we examined in detail the F814W images from the 
MDS database for {\em all} of the galaxies in $z = 0.9$ cluster 
CL1603$+$4304, which is one of the target fields included in this 
work (u2845).  Most of the cluster members are much brighter than 
the EROs.  Since the cluster is quite distant, the cluster members 
have similar apparent sizes and suffer from similar cosmological 
dimming as the EROs.  Morphological classifications for these galaxies 
were published in ~\citet{Lubin98}.  Working down the full range 
of apparent magnitudes, it was increasingly difficult to classify 
the fainter galaxies into the traditional galaxy morphological 
types (Elliptical, S0, Sa/Sb/Sc Spirals, Irregulars), although it 
was still possible to distinguish between spheroidal systems 
(presumably mostly elliptical and S0 galaxies) and disky systems 
(Sb/Sc/Irr).  Spirals of type Sa represent a somewhat fuzzy boundary.  
Depending on the SNR or strength of any ongoing star formation, 
these could be classified either way.  We therefore opted to simplify 
our morphological classification of the EROs.

We visually classified the galaxy morphologies into four broad 
categories.  These categories were selected to parallel the MDS 
classifications, as well as reflect the difficulties in placing 
faint, high-redshift sources into the traditional morphological 
classes, however there are close parallels between the two systems.  
EROs classified as spheroids or pure bulge galaxies (B) show no 
convincing evidence for the presence of a disk.  Bulge-dominated 
(BD) systems show evidence of a disk, but the majority of the 
luminosity is coming from the spheroidal component, and any disk 
component is generally featureless.  Disky systems with some evidence 
of a bulge (DB) generally show some evidence for structure in the 
disk (e.g. spiral arms or dust features), and the luminosity is not 
generally dominated by the spheroidal component.  Finally, disks (D) 
do not show any obvious bulge component, and often have mottling or 
other structure visible in the disk.  Some sources were too low of a 
SNR to classify, so they were listed as unclassifiable (U).  We 
further noted whether the ERO appeared to be undergoing a strong 
merger or interaction with other nearby galaxies regardless of their 
color.  

All of the EROs in our HST images were classified independently on 
a video display by both authors.  We then combined the classifications 
and reviewed the sources together to resolve the disagreements, which 
involved $\sim$20\% of the sample, and to settle on a final 
classification.  We also reviewed the EROs grouped into their separate 
categories.  This review involved looking at our reprocessed images as 
well as the MDS pipeline-processed images.  

\subsection{MDS Profile Fitting}

For a more quantitative analysis of the galaxy properties, we utilized 
the galaxy profile fitting results from the MDS database.  For almost 
all of our EROs, the MDS database includes a set of morphological 
parameters derived from their automated object detection and 
classification algorithms.  More detailed information on the entire
MDS reduction pipeline and morphological classification process can be 
found in ~\citet{rat94} and ~\citet{rat99}, as well on the MDS
website at \url{http://astro.phys.cmu.edu/mds.html}.

The MDS automated object classification involved a two-dimensional maximum
likelihood estimator (MLE) analysis that automatically optimizes the model
and the number of parameters to be fitted to each object image. 
Two scale-free, axisymmetric models are chosen to describe the galaxy 
profile. The spheroidal component, which would include elliptical galaxies 
and the bulges of spirals, is assumed to follow a de Vaucouleurs profile, 
while the extended disk component follows an exponential profile.
Each profile is characterized by a major axis half-light radius and axis
ratio. Point-like stellar sources are examined through the same procedure,
except that a Gaussian profile is used. A maximum likelihood parameter
estimation is used to determine the best model and the parameter values. 
For each set of model parameters, a model image of the object is created
and compared with the actual object images. Finally, a best-fit model
and its parameters are classified with the following categories for 
resolved sources: bulge, bulge$+$disk, disk, and galaxy (generally for 
low SNR sources where neither a bulge nor disk classification is 
significant).  We use the model bulge to total luminosity ratio to divide 
bulge$+$disk galaxies into our classes BD (for $L_B/L_{Total} \geq 0.5$) 
and DB.  This method does not classify irregulars, mergers, or 
interacting systems as such.

\subsection{Comparison of Classifications}

\begin{figure}[!ht]
   \plotone{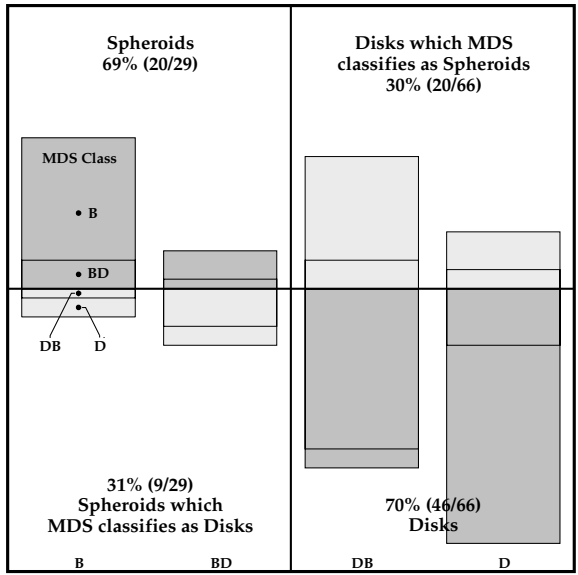}
   \caption{Cross comparison of the visual and MDS morphological
   classification results.  The darker gray areas are where the two
   methods agree well.  Overall, there is a 70\% agreement in the
   broad division of EROs into ``spheroids'' or ``disks.''
   \label{crossclass}}
\end{figure}

Figure~\ref{crossclass} shows a comparison of the visual and MDS 
classification results for EROs classified by both methods.  We 
group the B and BD classes under the label ``spheroids'' and the 
DB and D classes under the label ``disks.''  The darker gray areas 
show where the two methods agree reasonably well, while the lighter 
gray areas show the disagreements.  Overall, the agreement is good, 
where both the ``spheroids'' and ``disks.'' in the two classification 
methods agree at the 70\% level.  The largest source of disagreement 
comes from EROs which were visually classified as DB (disk with a 
bulge), but which the MDS classified as a pure bulge.  A majority of 
these systems are relatively faint in the HST images, where it can 
be difficult to distinguish between an extended spheroid or a disk, 
but also acknowledge the possibility of a fundamental failure in 
properly classifying sources visually.  Most of the remaining 
disagreement stems from the split between BD and DB, and can be 
tuned out by modifying the division in the MDS bulge-to-total 
luminosity ratio from the arbitrary 50\% bulge-to-total adopted 
for this study.

\section{Results \& Discussion}

\subsection{Morphological Distribution}

Our morphologies are based on F814W images.  Assuming the median 
redshift of one for EROs from the ~\citet{cim02} sample with 
K$ \leq 19.2$, the WFPC2 data sample a rest-frame wavelength of 
4100\,\AA.  The F814W images thus represent a compromise between 
sensitivity to star formation at shorter restframe wavelengths and 
better probing any extended old stellar populations at longer 
restframe wavelengths.  As shown in simulations by ~\citet{hibvac97}, 
morphological classifications using F814W images do not show any 
significant biases at $z \sim 1$.

\begin{figure}[!ht]
\plotone{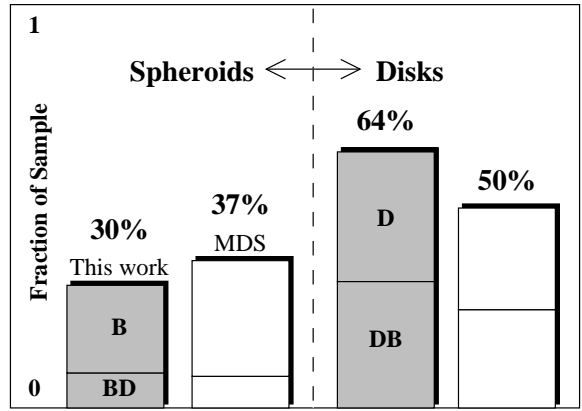}
\caption{Relative fraction of spheroidal galaxies vs. disks as classified
in this survey (gray bars) and by the MDS automated profile fitting (white
bars).  Not included in this plot are the 6\% of sources unclassifiable
by us or the 14\% unclassified by the MDS.  \label{relfrac}}
\end{figure}

Using the results from our visual classification, we find that 
30$\pm$5\% of our EROs have morphologies consistent with spheroidal 
(B and BD) galaxies. Disks (D and DB) dominate the EROs at 64$\pm$7\%\ 
of the sample.  Only 6\% of the EROs were unclassifiable, due primarily 
to low SNR on the WFPC images.  The uncertainties are derived simply 
from the square root of the number of EROs in that subset, and does not
try to include the unclassifiable sources.  We plot in Figure~\ref{relfrac} 
the relative fractions of spheroids and disks in our visual classification 
as well as from the MDS profile fitting.  

We find that the relative morphological mix of EROs in subsets constructed 
of the EROs in known foreground galaxy cluster fields vs. those selected in 
the remaining ``blank'' fields is consistent to within the uncertainties.
Since any lensing should be unbiased with respect to background galaxy 
morphologies, we do not differentiate between cluster and field sources in 
the remaining analysis.

Some of the EROs, including both disks and spheroids, appear to be 
involved in recent mergers or show evidence of strong interactions 
(e.g. tidal tails, strong asymmetries).  These comprise 17$\pm$4\% 
of the sample.  Without additional information, these systems represent 
the most likely source of possible large-scale starbursts.  
This is consistent with the observation that among {\it bright} EROs, roughly 
20\% have 850$\mu$m detections ~\citep{and99}.
One third of the galaxies we classed as spheroids have one or more faint 
companions or show signs of recent interaction, suggesting that a significant 
fraction of otherwise old stellar populations may not be in purely 
passively evolving systems.  

The morphological classifications derived from the MDS profile fitting
are largely consistent with our visual results, but show a slightly higher 
fraction of spheroidal galaxies, with 37$\pm$6\% of the sample classed as 
B or BD (with a bulge fraction larger than 50\%).  There is a corresponding 
drop in the disk fraction, with 50$\pm$7\% classed as DB or D, but disks 
still dominate the overall $\rm (I-K)\geq 4$ ERO population.  A larger 
fraction of the sample, 14$\pm$3\%, were not fit with the bulge$+$disk 
models due to low SNR or a more conservative avoidance of the CCD edges 
in the WFPC2 data.

We find that our ERO sample selected with $(F814W - Ks) \geq 4$  
is dominated by disk galaxies, and not by spheroids or strongly 
interacting systems.  
The star formation rates in these disks could span a wide range, 
including normal galaxies with fairly quiescent star formation.  HR\,10 
type systems, with mostly young stars and undergoing massive starbursts, 
may comprise only a small fraction of the sample.  The 
origins of their red colors may be traced to a significant 
old stellar population combined with some dust extinction, especially 
considering the edge-on orientation of many of the disks in our sample 
of EROs.  Even though most of the stellar mass may already be in place 
by $ z \sim 1 - 2$ for these EROs, such a large fraction of disk 
galaxies implies that there is still a substantial amount of gas 
available to feed on-going star formation.  

The existence of such a large fraction of disk galaxies and interacting 
systems in our sample suggests that hierarchical merging may be an 
important mechanism for the formation and evolution of the ERO population.
However, the scenario in which ellipticals were formed in a ``monolithic 
collapse'' at high redshifts and evolve passively thereafter cannot be 
excluded.  While 30\% of our ERO sample are clearly 
early type galaxies, whose colors are consistent with
old stellar populations formed at high redshifts. 
We also found that one-third of the spheroids in our sample
have faint companions or signs of interaction. This suggests 
that although a majority of their mass could
be assembled rapidly at high redshift, these systems are
not simply isolated, passively evolving old stellar
populations, and continuing accretion of gas or merger
events plays a significant role in their evolution. Examples
of secondary star formation in field E/S0s ($z \sim 0.1 - 0.73$) 
can be found in ~\citet{treu02}.

Our results contradict those of ~\citet{mor00}, which are also
based on HST morphologies.  They find that 50--80\%\ of their sample 
have E/S0 morphologies on the basis of one-component exponential model 
fits.  However, we note that their ERO sample was assembled from the 
published literature, with the corresponding heterogeneous selection 
functions of the original surveys.  In addition, the morphologies 
were determined on HST images from both WFPC2 and NICMOS, probing widely 
different rest-frame wavelengths and thus differing sensitivities to 
star formation or old stellar population.

\subsection{Surface Density}
\label{SDsect}

\begin{figure}[!ht]
\plotone{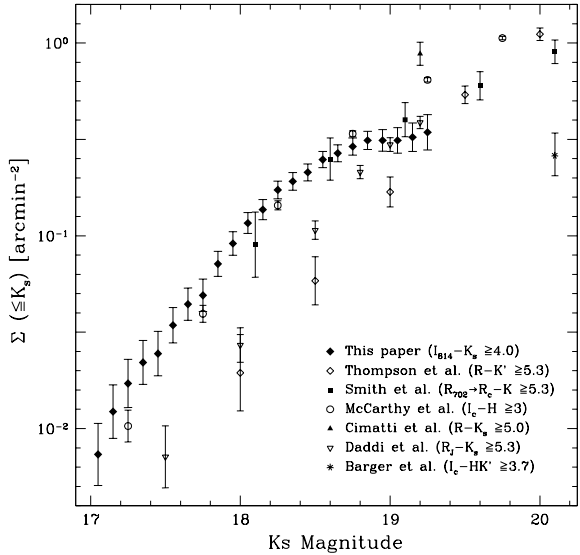}
\caption{Cumulative ERO surface density from this work (filled diamonds), 
as well as several other surveys for comparison.  The uncertainties are 
derived assuming only Poisson counting statistics in each bin.  
\label{surfdens}}
\end{figure}

The integrated surface density of EROs, is defined as total number 
of EROs brighter than a given magnitude per unit area on the sky.  
Because the area covered in this survey is a function of the magnitude, 
the differential number of EROs selected in each magnitude bin is a 
function of both the magnitude and the area surveyed.  Calculating the 
integrated surface density thus required rescaling the number of EROs 
in each of the brighter bins by the appropriate area ratio prior to 
integration.  We plot the resulting surface density of EROs from our 
survey derived by this method in Figure~\ref{surfdens} (filled diamonds).  
The uncertainties were derived simply from the square root of the rescaled 
number of EROs.  

For comparison, we also plot in Figure~\ref{surfdens} the results from 
other recent surveys ~\citep{tho99,smith02a,daddi00a,LCIR01,cim02,barg99}.  
The primary difficulty in making such comparisons is that each survey 
used a different set of filters and different selection criteria for 
identifying EROs.  In order to make a general comparison between these 
disparate surveys, we make several simplifying assumptions.  First, we 
treat all K filters as functionally equivalent (e.g. $K == K^\prime == K_s$), 
so no color terms are applied to convert magnitudes between them.  The same 
is true for the several different I filters and R filters used.  We note 
that these assumptions are generally made in most recent published ERO 
surveys unless the color selection limit is explicitly tied to some 
fiducial model SED, typically a passively evolving $z=1$ old stellar 
population (i.e. elliptical galaxies), {\em and} the specific filters 
used for the survey.  Second, we adopt the generic colors of 
($R-I$)$=$1\fm3 and ($H-K$)$=$1\fm0 to convert the surveys based on R 
band or H band data to our $I-K$ colors.  Third, we plot without 
additional correction the ~\citet{cim02} results, even though they 
select EROs at a bluer limit: ($R-K_s$)$\geq$5\fm0.  Finally, we adopt 
the ~\citet{barg99} results directly for ($I_c-HK^\prime$)$\geq$3\fm7, 
which they show to be equivalent to ($I-K$)$\geq$4\fm0 for that filter 
set.

While the generic color conversions we adopt obviate the possibility 
of more detailed comparisons between the different ERO surveys, they 
do allow us to consider broad trends in the surface densities as a 
function of the selection filters.  First, the two I-band based 
surveys (this work and ~\citet{LCIR01}) agree quite well over the 
region $17^{\rm m} \leq K_s \leq 19^{\rm m}$.  This suggests that our 
sample is neither significantly inhomogeneous nor incomplete over this 
range, although the turnover in counts to fainter magnitudes suggests 
incompleteness in our sample for $K>19^{\rm m}$.  The two larger R-band 
based surveys ~\citep{tho99,daddi00a} also agree with each other over 
this same range in K-band magnitude, but are systematically about a 
factor of three lower than the I-band surveys.  The lensing-corrected 
surface density of EROs from the cluster-pointed survey of 
~\citet{smith02a}, despite using an R-band filter, agrees well with 
the I-band based surveys at brighter magnitudes ($K<19^{\rm m}$).  
The other two surveys, ~\citet{cim02,barg99}, differ from the other 
results.   

Aside from the uncertainties in converting from one filter system to 
another, there are two primary effects which can qualitatively account 
for the similarities and differences in the surface densities of EROs 
from these different surveys: cosmic variance, and color selection 
effects.  Both the ~\citet{cim02} and ~\citet{barg99} surveys cover 
relatively small, connected areas on the sky, and are thus more subject 
to cosmic variance, especially considering the strong clustering seen 
in ~\citet{daddi00a}.  The ~\citet{smith02a} survey is composed of 10 
widely-separated sight lines and thus should be less sensitive to cosmic 
variance, but their sample does show a wide field-to-field variation in 
the number of EROs.  ~\citet{cim02} selects EROs at a bluer limit, 
($R-K_s$)$\geq$5\fm0, which likely contributes to their higher ERO 
surface density.

A color selection effect may contribute to the apparent differences in 
surface density between R-band based ~\citep{tho99,daddi00a} 
and I-band based (this work and ~\citet{LCIR01}) ERO surveys.  We 
offer below (see \S\ref{ColSel}) a qualitative argument on this, as 
we do not have a proper multiband deep survey to address this with 
real data.  

\subsection{Volume Density}

We can make an estimate of the volume density of EROs with some 
assumptions on the range of redshifts at which they may be found, 
and compare these results with the local density of massive galaxies.  
The color selection limit of ($I-K$)\,$\geq$\,4\fm0 sets the lower 
redshift bound to $z = 1$, which is appropriate for passively evolving 
old stellar populations.  However, photometric uncertainties could make 
this $z = 1$ boundary somewhat fuzzy.  Galaxies with significant dust 
extinction (see the following section) could also lie at lower redshifts.
\citet{cim02} obtained spectroscopic redshifts for a sample of EROs 
with ($R-K$)$>$5 which includes sources of both types down to $z \sim 0.7$.  
We adopt an upper cutoff to the assumed redshift range of $z = 2$, as 
higher redshift EROs would be anomalously luminous given their bright 
$K$-band magnitudes.  

Under the above assumptions, we derive a co-moving volume covered by 
our survey to be $4\times10^5\,h_{70}^{-3}$\,Mpc$^3$.  This volume is
only weakly dependent on the assumed redshift range, and only changes 
by a factor of two if the redshift range is narrowed to $z \sim 1.0 - 
1.5$ or broadened to $z \sim 0.7 - 2.8$.  This volume was derived from 
the total survey area of 409 square arcminutes, and does not take into 
account the variable survey depth with magnitude. 

The galaxies classed as spheroids in our survey have a co-moving 
volume density of $1\times10^{-4}\,h_{70}^3$\,Mpc$^{-3}$.  
The disks have a co-moving volume density about twice as large, 
and the total ERO sample (115 EROs) reach a density of 
$3\times10^{-4}\,h_{70}^3$\,Mpc$^{-3}$.  

To compare with nearby massive galaxies, we adopt the local K-band
luminosity functions for early-type and late type galaxies from
~\citet{Kochanek01}.  We integrate from 10L$^*$ down to 1L$^*$,
which corresponds to our K-band limit at $z = 1$ after correcting
for passive evolution and cosmological K-corrections.  We find that the
EROs can account for only one-third of the local massive galaxies,
and that the relative morphological mix is about the same in the two
samples.  This is reasonable, but should be considered only an 
order-of-magnitude agreement given the factors of $\sim$2 uncertainties 
in the volume densities arising from the assumptions on the redshift 
distributions, area surveyed as a function of depth (\S\ref{sec:area}), 
and contamination (\S\ref{sec:eodisks}).

\subsection{Dust Extinction in Disky EROs}
\label{sec:eodisks}

\begin{figure}[!ht]
\plotone{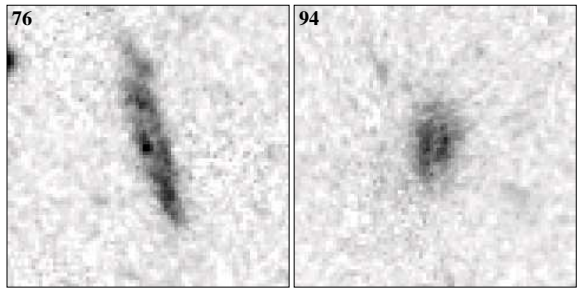}
\caption{Examples of two edge-on disks where dust extinction may contribute
significantly to their overall ($I-K_s$) color.  Each image is 8'' square.
The labels correspond to their source numbers from Table~\ref{ERO_tab}.
\label{edgeon}}
\end{figure}

While classifying the EROs, we also noted that 40\% of the disky 
systems (DB and D) appeared to be sufficiently edge-on that even small 
amounts of dust in a disk could have a disproportionately large effect 
on the overall system color.  These systems are noted in Table~\ref{ERO_tab} 
with italicized entries under morphology: {\em DB} and {\em D}, and we show 
two examples of edge-on EROs in Figure~\ref{edgeon}.  This is far more 
than expected from a set of randomly oriented galaxies, suggesting that 
orientation effects are responsible for their inclusion in the ERO sample.

Given the potentially large extinctions possible from dust in otherwise 
normal disk galaxies, it is possible that the edge-on systems are at 
lower redshift ($z < 1$).  Several of these show extended disks of 
large apparent size (several arcseconds), which would be unusually large
(tens of kiloparsecs) if at $z \sim 1$ or more.  The edge-on systems 
comprise half of the disky EROs, or one-third of the total ($I-K$) 
selected ERO sample.  They thus represent a large and previously 
unanticipated source of contamination in the ERO population.  

\subsection{Color Selection Effects}
\label{ColSel}

How comparable are ERO samples selected using an $(R - K) \ge 5.3$ 
color versus an $(I - K) \ge 4$ color?  This important issue has 
never been clearly addressed.  We investigated this issue using model 
spectral energy distributions, but lack the necessary multiband data 
to compare to the models.  Deep, wide-field infrared/optical surveys 
should be able to address this point in more detail.  

\begin{figure}[!th]
\plotone{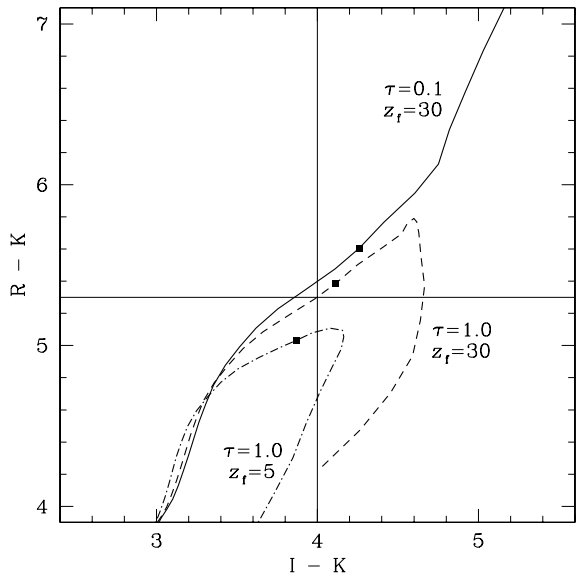}
\caption{$RIK$ colors for a ~\citet{BC96} model
approximating an old stellar population (OSP, $\tau = 0.1, z_f=30$),
as well as two models with longer exponential decay times ($\tau = 1.0$)
but differing formation redshifts ($z_f = 30, 5$).  All three curves are
marked with squares at $z \sim 1$.  Note that that pure, passively-evolving
OSPs would be selected in either an ($R-K$) or ($I-K$) survey for EROs, but
that any residual star formation, such as from a disk, would quickly drop
objects out of the ($R-K$) sample.  Reducing the redshift of formation much
below five, or increasing the residual star formation rates ($\tau > 1$)
would drop objects out of both samples.  Dust counteracts these effects
somewhat, shifting objects towards the upper right of this plot.
\label{modelRIK}}
\end{figure}

\begin{figure}[!ht]
   \plotone{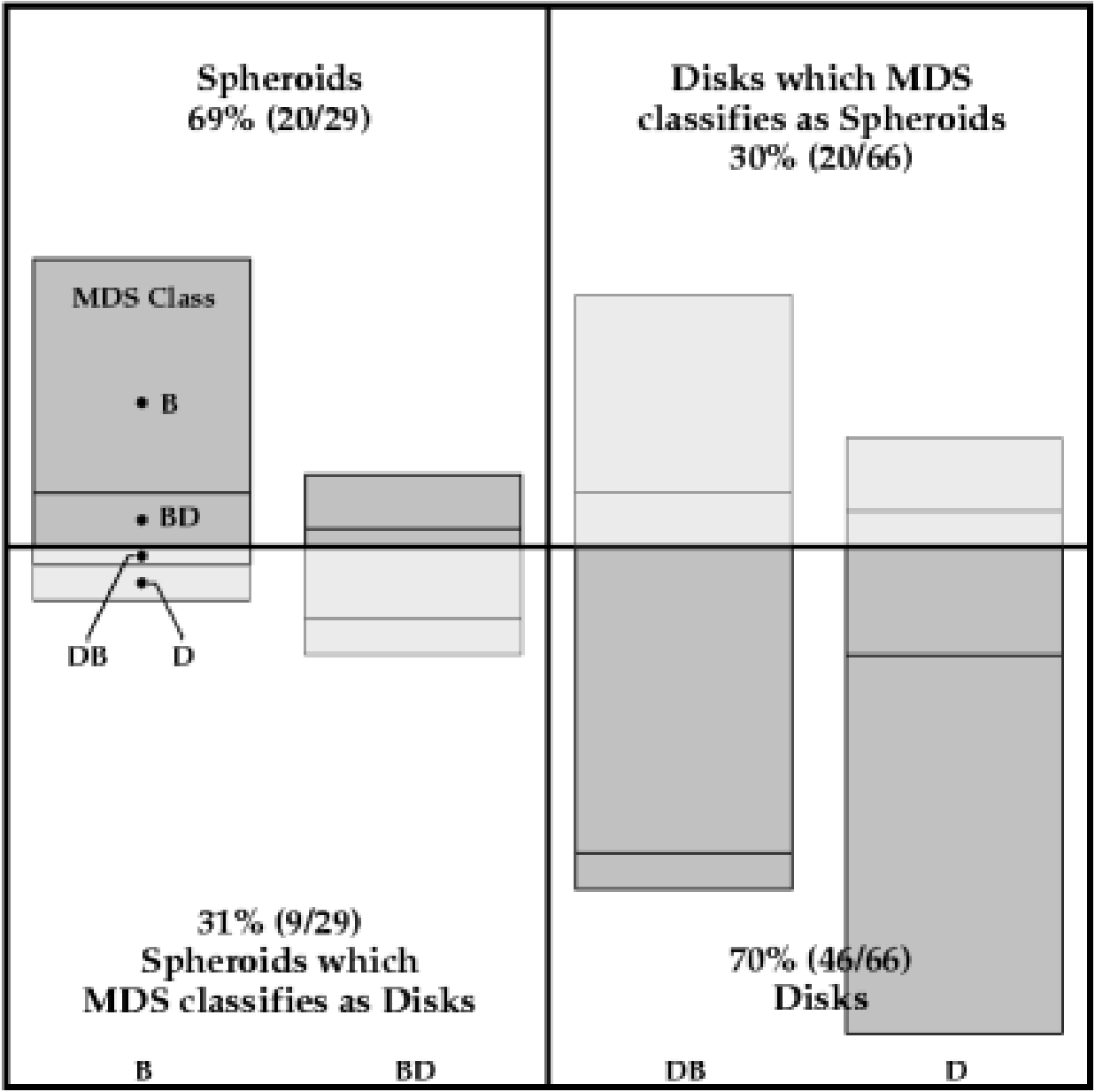}
   \caption{Full-resolution HST/WFPC2 F814W images of all 115 EROs.  Each
   image is 8\arcsec square, rotated from the default HST orientation to
   have north up and east to the left.  The numbers in each subimage
   correspond to the object numbers in Table~\ref{ERO_tab}.  \label{hstimgs}}
\end{figure}

In Figure~\ref{modelRIK} we plot the ($R-K$) vs. ($I-K$) colors for 
a ~\citet{BC96} model approximating an old stellar population (OSP, 
$\tau = 0.1, z_f=30$) or passively evolving elliptical galaxy.  We also 
plot two models with longer exponential decay times ($\tau = 1.0$) but 
differing formation redshifts ($z_f = 30, 5$), which should contain 
a significant fraction of old stars in the range of $1 < z < 2$, but 
still have some residual star formation.  A similar plot covering the 
$VIH$ color-color plane can be found in ~\citep[their Figure 2]{LCIR01}.  
Their data show that EROs selected with ($I - H$)\,$\geq$\,3 have a wide 
scatter in the $(V - I)$ color, which the authors interpreted as due to 
prolonged star formation.

The two dotted lines in Figure~\ref{modelRIK} mark the fiducial colors 
($R - K$)\,$=$\,5\fm3 and ($I - K$)\,$=$\,4\fm0, representing the 
colors of a $z = 1$ passively-evolving elliptical galaxy used by most 
surveys to select EROs.  The squares mark the $z \sim 1$ points in each 
model curve.  Assumptions on the models used, as well as the assumed 
cosmology and the specific filter bandpasses used for ERO surveys 
can account for several tenths of a magnitude variation in the 
expected colors of a $z = 1$ passively-evolving elliptical galaxy.

Several predictions can be made from this color selection effect.  First, 
is that both R-band and I-band based ERO surveys should select the {\em 
same population} of passively-evolving old stellar populations 
(elliptical galaxies).  Second, I-band based ERO surveys should 
preferentially include disk galaxies.  Light from a bulge comprised of 
older stars would dominate the ($I-K$) color, while even small amounts 
of residual star formation in the disk keeps them too blue in ($R-I$) 
to be included in an ($R-K$) selected sample.  Other factors, such 
as dust extinction, may counteract the star-formation and contribute 
significantly to their overall color.  This implies that ERO samples 
selected on their ($R - K$) color may not be comparable to samples 
selected on their ($I - K$) color.

Figure~\ref{modelRIK} can also be used to set some constraints on the 
formation redshift for the EROs.  Models with exponential decay times 
longer than about 1.5\,Gyr ($\tau \geq 1.5$), or with a formation 
redshift lower than five ($z \leq 5$) simply have too much residual 
star formation.  Without any reddening from dust, the blue light from 
a young population of stars would be sufficient to drop these galaxies 
out of either  ($R-K$) or ($I-K$) samples of EROs.  Thus, the brighter 
EROs classed as spheroids, especially those without any evidence of 
a disk or ongoing star formation, are likely to have formed a majority 
of their stars at relatively high redshift ($z > 5$).

The morphological mix in our sample of relatively bright, ($I-K$)-selected 
EROs is similar to that of ~\citet{smith02a} among a sample of fainter 
EROs ($K \leq 20.6$, $R-K \geq 5.3$) identified in the fields of foreground 
clusters massive enough to gravitationally lens the higher redshift EROs.  
They classify 18\% of their sample as compact, and 50\% as irregulars 
(including disk-like systems), while 32\% are too faint to be classified. 

Considering redder subsamples from both this work ($F814W - K_s \geq 5$) 
and ~\citet{smith02a} ($R - K \geq 6.0$), we again find similar results.
Of the 11 redder EROs in our sample, nine were classifiable.  Of these 
nine, 89$\pm$31\% have disk morphologies, a much higher fraction than 
in the full sample.  The redder ~\citet{smith02a} subsample has $\sim$90\%
with disk/irregular morphologies.  

The expectation from the color selection effect is that (R-K) samples 
should contain a higher fraction of spheroids at a given $K$ magnitude 
limit.  However, the fainter ~\citet{smith02a} sample, about three magnitudes 
deeper than our ERO sample, is composed primarily of irregulars and 
disks.  This suggests that the morphological mix of EROs does change at 
fainter magnitudes, and ~\citet{smith02a} conclusion that fainter and 
redder samples are dominated by massive, dusty starbursts does not 
contradict our findings.  Clearly, this color selection effect needs to 
be investigated further, with larger and deeper samples of EROs with both 
(R-K) and (I-K) selection on the same area of sky, so that a proper 
comparison can be made.  

\section{Summary}

Our results highlight the complex nature of optical/near-IR color
selected EROs.  The high resolution HST morphologies indicate that 
disks are the dominant constituent of the bright ($K < 20$\,mag) ERO 
population selected with an $(I - K)$ color greater than or equal to 
4 magnitudes.  Galaxies classified as spheroids, which can be 
considered comparable to passively evolving elliptical or S0 galaxies, 
and strongly interacting systems which may represent dusty starbursts, 
only contribute small fractions to the total ERO population.  There 
are real differences in the surface densities of EROs selected by their 
$(R - K)$ color as compared to the $(I - K)$ samples, which may 
reflect a preferential selection of disks in I-band based ERO surveys 
and in any case complicates the comparison of various surveys and 
their interpretation.  In addition, edge-on disk galaxies comprise 
a significant fraction of our ERO sample: 40\% of the disks, or 
28\% of the full sample.  Even small amounts of dust in this 
orientation could redden otherwise normal disk galaxies at lower 
redshift ($z < 1$) sufficiently to be included in the ERO samples.  
Our results imply that hierarchical merging and continuing accretion of
gas still play an important and continuing role in the evolution of massive
galaxies, even though most of their stellar mass may already be in place
by $z \sim 1$.

\acknowledgements
We thank Myungshin Im for helpful discussions about the MDS.
This research was supported by the HST grant AR-08756 and the 
NASA Long Term Space Astrophysics program under the grant NAG5-10955. 
The MDS is based on observations with the NASA/ESA Hubble Space Telescope, 
obtained at the Space Telescope Science Institute, which is operated by 
the Association of Universities for Research in Astronomy, Inc., 
under NASA contract NAS5-26555.



\clearpage

\begin{deluxetable}{lrrrrr|lrrrrr}
   \tabletypesize{\scriptsize}
   \tablecaption{Observations. \label{obs_tab}}
   \tablewidth{0pt}
   \tablehead{
      \colhead{MDS}                    &
      \colhead{N$_{\rm ERO}$}          &
      \colhead{$5\sigma\ K_s$}         &
      \colhead{FWHM}                   &
      \multicolumn{2}{c}{Exp Time (s)} &
      \colhead{MDS}                    &
      \colhead{N$_{\rm ERO}$}          &
      \colhead{$5\sigma\ K_s$}         &
      \colhead{FWHM}                   &
      \multicolumn{2}{c}{Exp Time (s)} \\
      \colhead{Field}                  &
      \colhead{}                       &
      \colhead{limit\tablenotemark{a}} &
      \colhead{arcsec}                 &
      \colhead{$K_s$}                  &
      \colhead{$F814W$}                &
      \colhead{Field}                  &
      \colhead{}                       &
      \colhead{limit\tablenotemark{a}} &
      \colhead{arcsec}                 &
      \colhead{$K_s$}                  &
      \colhead{$F814W$}                }
   \startdata
      u2845i8 &  5 & 18.85 & 1.25 &  5400 & 16000 & ufj00i2 &  3 & 19.02 & 1.37 &  4800 &  4200 \\
      u29g1i6 &  0 & 18.55 & 1.43 &  5400 &  6600 & ug502i4 &  0 & 18.54 & 1.68 &  3750 &  1700 \\
      u29g3i6 &  0 & 18.71 & 1.68 &  4350 &  6600 & ugi00i2 &  0 & 18.27 & 1.74 &  2700 &  4700 \\
      u2b15i6 &  2 & 18.62 & 1.62 &  4500 &  3600 & ugk00i2 &  0 & 18.45 & 1.56 &  3750 &  5400 \\
      u2c41i8 &  0 & 18.85 & 1.56 &  6450 & 16800 & uha01i3 &  0 & 18.97 & 1.68 &  6750 &  4200 \\
      u2c47i6 &  3 & 19.29 & 1.50 &  9000 & 12600 & uhg00i2 &  1 & 19.25 & 1.62 &  9750 &  5600 \\
      u2c48i6 &  3 & 18.63 & 1.99 &  4500 & 12600 & uho00i3 &  1 & 18.89 & 1.62 &  4500 &  4500 \\
      u2fl1i5 &  2 & 18.77 & 1.68 &  4500 & 10500 & uih00i2 &  0 & 18.58 & 1.87 &  4500 &  4200 \\
      u2fq1i5 & 14 & 18.67 & 1.25 &  4800 & 10500 & uim03i3 &  4 & 19.07 & 1.56 &  7650 &  4000 \\
      u2fq2i6 &  1 & 18.58 & 1.37 &  5400 & 12600 & ujh01i2 &  1 & 18.87 & 1.99 &  4500 &  4200 \\
      u2gk1i3 &  1 & 18.62 & 1.68 &  4500 &  4800 & uko01i2 &  2 & 19.07 & 1.56 &  5850 &  4200 \\
      u2h91ic &  2 & 18.80 & 1.37 &  5400 & 28800 & ulj00i4 &  3 & 19.49 & 1.43 & 13050 &  5100 \\
      u2iy1i6 &  1 & 18.43 & 1.93 &  3750 &  6400 & uo501i3 &  5 & 18.93 & 1.50 & 12450 &  6300 \\
      u2iy2i6 &  0 & 18.67 & 1.62 &  4500 &  6400 & upj00i2 &  0 & 18.97 & 1.56 &  4500 &  4200 \\
      u2uj2i3 &  1 & 18.64 & 1.43 &  4350 &  3600 & uqc00i2 &  2 & 18.64 & 1.25 &  5400 &  4200 \\
      u2uj7i3 &  2 & 18.71 & 1.68 &  3750 &  3600 & uqc01i2 &  2 & 19.15 & 1.43 &  9150 &  7200 \\
      u2um1ia &  0 & 18.72 & 1.37 &  3750 & 11000 & uqg00i2 &  0 & 18.42 & 1.31 &  2700 &  3600 \\
      u2v12i5 &  0 & 18.22 & 1.99 &  3000 &  6700 & uqj10i3 &  3 & 18.93 & 1.50 &  5250 &  4100 \\
      u2v14i5 &  0 & 18.62 & 1.25 &  4200 &  6700 & uqk00i2 &  6 & 19.26 & 1.50 &  9000 &  2450 \\
      u2v15i5 &  0 & 18.34 & 1.62 &  2700 &  6700 & uqk02i4 &  2 & 18.89 & 1.62 &  4500 &  6600 \\
      u2v16i5 &  1 & 19.07 & 1.56 & 10650 &  6700 & uqk04i4 &  1 & 19.06 & 1.56 &  4500 &  4000 \\
      u2v18i5 &  1 & 18.72 & 1.87 &  4350 &  6700 & uqk11i5 &  3 & 18.93 & 1.37 &  4350 &  3100 \\
      u2v19i5 &  2 & 18.73 & 1.99 &  4200 &  6700 & uql00i2 &  1 & 18.78 & 2.12 &  6000 &  4200 \\
      u3063i6 &  1 & 18.95 & 1.37 &  4500 & 14400 & uri01i3 &  0 & 18.59 & 1.99 &  4500 &  3000 \\
      u30h1i4 &  0 & 18.75 & 1.74 &  4500 &  5000 & usa00i3 &  1 & 18.68 & 1.68 &  4500 &  6300 \\
      u30h2i4 &  0 & 19.07 & 1.56 &  4500 &  5300 & usa02i3 &  0 & 18.62 & 1.37 &  4050 &  6300 \\
      ubb10i2 &  0 & 18.45 & 1.93 &  4200 &  5800 & usc10i7 &  1 & 18.82 & 1.68 &  9050 &  4500 \\
      ubi02i2 &  3 & 19.30 & 1.68 &  8850 &  4700 & usc12i4 &  2 & 18.58 & 1.31 &  5250 &  4135 \\
      ubm00i3 &  1 & 18.81 & 1.56 &  4500 &  5400 & utb11i3 &  0 & 18.39 & 1.62 &  2700 &  9100 \\
      uci10i4 &  2 & 18.86 & 1.37 &  4950 & 10800 & uub01i2 &  0 & 18.29 & 1.68 &  2700 &  2700 \\
      udh00i2 &  1 & 18.80 & 1.56 &  4500 &  3300 & uuc04i6 &  2 & 18.72 & 1.62 &  7200 &  3900 \\
      udm00i2 &  0 & 18.69 & 1.74 &  4500 &  3000 & uvm01i2 &  1 & 19.02 & 1.37 &  4500 &  4600 \\
      udm10i3 &  1 & 18.75 & 1.87 &  4500 &  5400 & uwp00i3 &  1 & 18.61 & 1.99 &  4500 &  8400 \\
      uec00i2 &  4 & 18.48 & 1.37 &  2700 &  3000 & ux400i4 &  2 & 18.69 & 1.50 &  4350 &  7500 \\
      ued01i2 &  0 & 18.53 & 1.37 &  2700 &  5600 & uxn00i2 &  0 & 18.89 & 1.87 &  4500 &  2800 \\
      ueg00i3 &  0 & 18.60 & 1.37 &  2700 &  6300 & uxs10i3 &  1 & 18.74 & 1.43 &  4500 &  6200 \\
      ueh02i2 &  3 & 18.42 & 1.62 &  2700 &  4200 & uys00i2 &  1 & 18.61 & 1.87 &  4500 &  4600 \\
      uem00i5 &  0 & 18.69 & 2.06 &  4500 &  6600 & uzk03i3 &  1 & 18.84 & 1.81 &  4500 &  4900 \\
      ufg00i2 &  5 & 18.65 & 1.74 &  4050 &  4700 &         &    &       &      &       &       \\
   \enddata
   \tablenotetext{a}{The $K_s$ limiting magnitudes are derived from the
      sky noise in the center of each infrared mosaic image, and are quoted
      as 5$\sigma$ limits for a point source within an aperture diameter
      2.5 times the FWHM of the seeing.}
\end{deluxetable}

\clearpage

\begin{deluxetable}{rrrrrrrrl}
   \tabletypesize{\scriptsize}
   \tablecaption{Extremely Red Galaxies. \label{ERO_tab}}
   \tablewidth{0pt}
   \tablehead{
      \colhead{\#}                            &
      \colhead{MDS ID\#}                      &
      \colhead{Coordinates\tnm{a}}            &
      \colhead{$K_s$/$I-K_s$}                 &
      \colhead{$K\sigma$}                     &
      \colhead{$I\sigma$}                     &
      \multicolumn{2}{c}{Morph.Class\tnm{b}}  &
      \colhead{Comments\tnm{c}}               \\
      \colhead{}                              &
      \colhead{}                              &
      \colhead{$\alpha$ \hrulefill (J2000)
                        \hrulefill $\delta$}  &
      \colhead{}                              &
      \colhead{}                              &
      \colhead{}                              &
      \colhead{YT\tnm{d}}                     &
      \colhead{MDS}                           &
      \colhead{}                              }
   \startdata
        1 & ufg00\#140 &  0:18:22.2928 $+$16:20:54.975 & 18.05 / 5.85 &  6.8 &  7.3 & D        & G  & asym disk or 1fc         \\
        2 & ufg00\#070 &  0:18:29.1165 $+$16:20:56.226 & 18.27 / 5.23 &  7.2 &  5.8 & {\em DB} & B  & asym disk                \\ 
        3 & ufg00\#121 &  0:18:30.1373 $+$16:20:39.685 & 18.29 / 4.27 &  6.4 & 12.6 & D        & B  & faint asym disk?         \\
        4 & ufg00\#044 &  0:18:31.2524 $+$16:20:43.706 & 17.68 / 4.16 &  9.4 & 33.4 & B        & B  & mfc, merger?             \\
        5 & ufg00\#083 &  0:18:31.5131 $+$16:20:40.797 & 18.05 / 4.60 &  5.9 & 15.5 & B        & G  & 1fc + tail, merger?      \\
        6 & uhg00\#049 &  0:20:11.5736 $+$28:36:51.713 & 18.12 / 4.24 &  9.6 & 26.6 & ID       & DB & DB + lg tidal tail       \\
        7 & udh00\#057 &  0:45:02.1324 $+$10:34:42.958 & 18.58 / 4.05 &  5.6 & 13.1 & D        & B  & faint disk               \\
        8 & ueh02\#120 &  0:53:35.3406 $+$12:49:29.408 & 18.28 / 4.15 &  5.8 & 16.5 & D        & D  & asym, LSB                \\
        9 & ueh02\#063 &  0:53:35.6556 $+$12:49:50.997 & 17.43 / 4.82 & 11.4 & 29.7 & B        & BD & faint disk               \\
       10 & ueh02\#104 &  0:53:39.9393 $+$12:49:34.931 & 18.08 / 4.92 &  5.8 & 15.2 & {\em DB} & B  & asym or 1fc              \\ 
       11 & ujh01\#118 &  1:09:03.3016 $+$35:35:36.262 & 18.71 / 4.62 &  5.4 & 11.7 & {\em D}  & D  & lg disk                  \\ 
       12 & ubi02\#055 &  1:09:56.7526 $-$02:26:18.601 & 18.49 / 4.03 &  8.6 & 23.6 & {\em D}  & B  & asym disk                \\ 
       13 & ubi02\#062 &  1:09:57.0151 $-$02:27:34.353 & 18.46 / 4.71 &  8.7 & 14.4 & B        & B  & faint disk?              \\
       14 & ubi02\#013 &  1:10:00.4624 $-$02:27:37.273 & 17.04 / 4.18 & 28.8 & 99.9 & B        & S  & stellar core + mfc, AGN? \\
       15 & uci10\#013 &  1:24:40.8825 $+$03:50:45.704 & 17.57 / 4.17 & 15.5 & 99.9 & B        & BD & bright spheroid          \\
       16 & uci10\#051 &  1:24:45.9369 $+$03:51:19.209 & 18.62 / 4.02 &  5.3 & 18.6 & ID       & D  & dbl nucl + mfc, merger   \\
       17 & ubm00\#093 &  2:01:50.2325 $-$11:41:14.232 & 18.29 / 4.30 &  7.6 & 27.3 & DB       & D  & asym disk                \\
       18 & ufj00\#077 &  2:07:01.3073 $+$15:26:18.414 & 18.67 / 4.08 &  5.0 & 23.1 & {\em DB} & DB & asym disk                \\ 
       19 & ufj00\#052 &  2:07:05.7003 $+$15:24:55.156 & 17.74 / 4.70 & 15.5 & 19.2 & {\em DB} & B  & lg LSB disk, asym nucl   \\ 
       20 & ufj00\#047 &  2:07:07.6300 $+$15:24:43.365 & 18.29 / 4.14 &  9.0 & 24.8 & B        & B  & stellar core + 2fc, AGN? \\
       21 & u2c48\#049 &  2:39:56.0682 $-$01:37:07.046 & 17.33 / 4.25 & 15.7 & 77.1 & ID       & D  & merger, lg tidal tail?   \\
       22 & u2c48\#078 &  2:39:59.2569 $-$01:37:21.172 & 17.55 / 4.35 & 13.2 & 46.7 & DB       & DB & lg disk, dust/arms       \\
       23 & u2c48\#114 &  2:40:00.5682 $-$01:37:07.237 & 18.22 / 4.45 &  6.3 & 28.1 & D        & DB & face-on LSB disk or mfc  \\
       24 & udm10\#100 &  2:42:52.0311 $-$00:05:08.190 & 18.45 / 4.73 &  5.0 &  7.2 & IU       & G  & spur or fc to disk       \\
       25 & ulj00\#053 &  2:43:50.2612 $+$37:17:53.956 & 18.49 / 4.34 & 11.7 & 15.0 & DB       & B  & faint asym disk or 1fc   \\
       26 & ulj00\#219 &  2:43:50.3033 $+$37:17:23.852 & 19.12 / 5.72 &  7.0 &  2.7 & {\em D}  & D  & v.faint LSB disk         \\
       27 & ulj00\#114 &  2:43:50.8212 $+$37:17:14.400 & 19.22 / 4.26 &  6.7 &  7.8 & D        & G  & irr, asym                \\
       28 & u2iy1\#022 &  3:02:47.3464 $+$00:13:08.655 & 17.22 / 4.45 &  9.8 & 53.8 & {\em DB} & DB & lg disk                  \\
       29 & u2v19\#079 &  3:38:37.9233 $-$00:13:03.154 & 18.33 / 4.30 &  6.5 & 17.6 & {\em D}  & D  & lg LSB disk              \\
       30 & u2v19\nd   &  3:38:38.5961 $-$00:12:42.727 & 17.69 / 4.15 & 12.4 & 48.9 & B        & -  & 2 EROs (2\arcsec sep)    \\
       31 & u2v18\#036 &  3:41:09.5269 $+$00:00:18.660 & 17.90 / 4.43 &  7.7 & 42.9 & B        & BD & asym                     \\
       32 & uim03\#102 &  3:55:31.1522 $+$09:44:47.822 & 18.60 / 4.04 &  7.7 & 15.0 & {\em DB} & D  & \nd                      \\
       33 & uim03\#089 &  3:55:32.3447 $+$09:44:49.683 & 18.57 / 4.02 &  6.6 & 18.9 & BD       & B  & faint disk?              \\
       34 & uim03\#100 &  3:55:32.8000 $+$09:44:47.262 & 18.34 / 4.51 &  7.6 & 12.7 & BD       & DB & asym                     \\
       35 & uim03\#075 &  3:55:35.4355 $+$09:42:43.514 & 17.94 / 4.63 &  8.1 & 16.5 & DB       & DB & asym                     \\
       36 & u2fl1\#038 &  4:14:41.8303 $+$05:35:47.750 & 17.84 / 4.50 &  7.2 & 39.7 & {\em D}  & DB & asym, 1fc                \\
       37 & u2fl1\#044 &  4:14:43.0621 $+$05:34:39.376 & 18.29 / 4.20 &  7.1 & 32.4 & B        & B  & 1fc                      \\
       38 & uho00\#069 &  4:16:55.7078 $-$05:59:36.262 & 18.64 / 4.62 &  6.0 &  9.9 & {\em DB} & DB & LSB disk                 \\
       39 & uko01\#043 &  4:56:49.0888 $+$03:52:37.939 & 18.21 / 4.10 &  6.7 & 34.5 & B        & D  & bright, resolved core    \\
       40 & uko01\#023 &  4:56:43.3215 $+$03:53:33.905 & 18.85 / 4.00 &  5.1 & 22.1 & B        & B  & 1fc                      \\
       41 & uqk11\#019 &  7:24:41.3649 $+$60:29:37.564 & 17.07 / 4.08 &  9.0 & 83.0 & D        & BD & lg disk, asym, dust      \\
       42 & uqk11\#077 &  7:24:43.8276 $+$60:31:37.111 & 18.21 / 4.39 &  6.2 & 17.9 & D        & D  & LSB disk or irr          \\
       43 & uqk11\#048 &  7:24:46.6626 $+$60:30:35.571 & 18.11 / 4.00 &  6.6 & 27.3 & {\em DB} & DB & asym                     \\
       44 & uqj10\#027 &  7:27:20.6173 $+$69:05:46.966 & 17.73 / 4.05 &  6.8 & 44.7 & {\em D}  & D  & LSB disk, 1bc            \\
       45 & uqj10\#062 &  7:27:25.4097 $+$69:06:17.115 & 18.42 / 4.65 &  7.4 & 10.9 & D        & DB & LSB disk or irr          \\
       46 & uqj10\#080 &  7:27:42.9072 $+$69:06:50.422 & 18.46 / 5.27 &  5.1 & 17.3 & {\em D}  & D  & asym                     \\
       47 & uqk02\#042 &  7:41:25.8953 $+$65:06:02.339 & 18.55 / 4.01 &  6.0 & 14.4 & DB       & DB & asym or 1fc              \\
       48 & uqk02\#064 &  7:41:31.6545 $+$65:06:09.883 & 18.51 / 4.40 &  6.3 & 11.3 & {\em D}  & D  & dbl nucl                 \\
       49 & uqk00\#133 &  7:42:37.6428 $+$65:06:32.076 & 18.38 / 5.38 &  7.1 &  4.9 & U        & G  & compact, asym            \\
       50 & uql00\#314 &  7:42:39.3423 $+$49:44:31.800 & 18.54 / 4.37 &  5.2 & 18.9 & U        & G  & v.faint, LSB             \\
       51 & uqk00\#299 &  7:42:44.2458 $+$65:05:49.770 & 18.77 / 4.36 &  5.8 &  9.7 & U        & D  & v.faint, asym            \\
       52 & uqk00\#096 &  7:42:44.6454 $+$65:05:49.863 & 18.71 / 4.12 &  6.3 & 12.0 & D        & D  & LSB disk or irr          \\
       53 & uqk00\#066 &  7:42:49.2476 $+$65:05:06.184 & 18.77 / 4.21 &  5.1 & 11.4 & D        & D  & LSB disk, asym or 1bc    \\
       54 & uqk04\#039 &  7:42:49.6767 $+$65:15:43.206 & 18.03 / 4.09 & 10.8 &  7.1 & B        & B  & asym nucl                \\
       55 & uqk00\nd   &  7:42:49.7006 $+$65:06:08.519 & 18.78 / 4.04 &  5.4 &  7.9 & B        & -  & asym                     \\
       56 & uqk00\#043 &  7:42:51.0075 $+$65:06:23.044 & 18.58 / 4.84 &  5.8 &  9.7 & B        & D  & asym                     \\
       57 & u2gk1\#190 &  8:30:48.3166 $+$65:51:12.636 & 18.22 / 4.99 &  6.1 &  4.7 & DB       & G  & asym, 1bc, merger?       \\
       58 & uvm01\#194 &  9:39:31.7625 $+$41:33:09.070 & 18.85 / 4.57 &  5.6 &  7.4 & {\em DB} & G  & asym                     \\
       59 & u2c47\#165 &  9:42:57.2880 $+$46:56:01.780 & 18.86 / 4.68 &  6.6 & 13.1 & D        & D  & asym, dbl nucl or dust lane \\
       60 & u2c47\#085 &  9:43:03.5351 $+$46:55:50.805 & 17.91 / 4.06 & 11.7 & 36.9 & BD       & DB & \nd                      \\
       61 & u2c47\#134 &  9:43:08.0085 $+$46:56:21.209 & 18.54 / 4.43 &  8.4 & 21.7 & {\em D}  & D  & asym, mfc                \\
       62 & uwp00\#051 & 10:02:26.2962 $+$28:50:00.903 & 18.03 / 4.79 &  7.2 & 22.6 & DB       & DB & asym, 2fc                \\
       63 & uxs10\#035 & 10:47:13.2802 $+$13:56:37.675 & 17.90 / 4.46 &  6.9 & 24.8 & DB       & B  & faint disk, 1fc          \\
       64 & uys00\nd   & 11:16:28.0883 $+$18:05:27.717 & 17.87 / 4.15 & 10.2 & 25.4 & B        & -  & stellar core, AGN?       \\
       65 & u3063\#511 & 11:40:29.4030 $+$66:07:58.555 & 18.86 / 6.05 &  5.2 &  2.6 & D        & D  & LSB disk or irr          \\
       66 & uzk03\#086 & 12:10:31.1231 $+$39:28:46.905 & 18.19 / 4.07 &  8.5 & 17.3 & DB       & B  & 1fc                      \\
       67 & u2b15\#029 & 13:33:35.2177 $+$16:50:11.271 & 18.33 / 4.07 &  5.4 & 34.5 & BD       & D  & faint disk, mfc          \\
       68 & u2b15\#046 & 13:33:37.1951 $+$16:50:00.346 & 18.32 / 4.01 &  6.0 & 17.6 & {\em D}  & B  & asym                     \\
       69 & u2uj2\#164 & 13:59:48.8443 $+$62:31:47.959 & 18.21 / 4.12 &  6.3 & 22.6 & DB       & DB & LSB disk                 \\
       70 & u2uj7\#082 & 13:59:54.7421 $+$62:28:35.463 & 18.14 / 4.20 &  6.5 &  9.6 & DB       & DB & \nd                      \\
       71 & u2uj7\#171 & 14:00:07.8212 $+$62:28:51.163 & 18.65 / 4.53 &  5.1 & 11.8 & U        & B  & compact, LSB             \\
       72 & ux400\nd   & 15:19:39.4365 $+$23:52:37.885 & 18.60 / 4.11 &  5.1 & 28.1 & {\em DB} & -\tnm{e} & lg asym disk + 1fc \\
       73 & ux400\#055 & 15:19:40.7129 $+$23:52:39.488 & 18.25 / 4.16 &  6.7 & 32.4 & BD       & DB & asym disk                \\
       74 & u2845\#097 & 16:04:15.4230 $+$43:04:15.840 & 18.75 / 4.56 &  5.1 & 20.0 & {\em DB} & B  & lg asym disk             \\
       75 & u2845\#077 & 16:04:17.4446 $+$43:03:47.643 & 18.70 / 4.64 &  5.3 & 33.4 & BD       & B  & 1fc?                     \\
       76 & u2845\#043 & 16:04:17.7243 $+$43:03:24.653 & 18.10 / 4.64 &  6.9 & 33.4 & {\em D}  & D  & lg LSB disk, dust lane   \\
       77 & u2845\#089 & 16:04:21.3952 $+$43:04:35.810 & 18.32 / 4.57 &  8.1 & 28.1 & {\em DB} & B  & \nd                      \\
       78 & u2845\#050 & 16:04:25.2398 $+$43:04:12.562 & 17.95 / 4.14 &  6.4 & 53.8 & B        & B  & 1bc                      \\
       79 & uuc04\#095 & 16:24:12.4485 $+$48:10:14.906 & 17.66 / 4.50 & 12.0 & 23.1 & DB       & DB & asym LSB disk, 1bc, merger? \\
       80 & uuc04\#044 & 16:24:15.3071 $+$48:09:37.509 & 17.89 / 4.14 &  6.2 & 32.4 & DB       & B  & mfc, tidal tail?         \\
       81 & usa00\nd   & 17:12:20.6910 $+$33:35:28.969 & 17.94 / 4.19 &  8.1 & 38.3 & ID       & -\tnm{e} & merger (poss barred sp?)\\
       82 & usc12\#072 & 17:22:37.0496 $+$50:13:35.231 & 17.94 / 4.13 &  7.9 & 28.1 & B        & B  & \nd                      \\
       83 & usc12\#045 & 17:22:37.6296 $+$50:13:00.220 & 17.88 / 4.05 &  9.8 & 34.5 & DB       & DB & lg late type spiral      \\
       84 & usc10\#056 & 17:23:00.3707 $+$50:10:54.617 & 18.22 / 4.02 &  5.1 & 24.2 & DB       & B  & 1fc, arm                 \\
       85 & uo501\#171 & 17:55:22.4072 $+$18:18:47.134 & 18.47 / 4.05 &  5.9 & 23.6 & B        & B  & \nd                      \\
       86 & uo501\#258 & 17:55:25.4940 $+$18:17:09.218 & 18.46 / 4.17 &  5.3 & 27.3 & {\em DB} & DB & asym, 2bc                \\
       87 & uo501\#207 & 17:55:26.3237 $+$18:17:15.018 & 18.51 / 4.52 &  6.1 & 22.1 & {\em BD} & D  & 1fc                      \\
       88 & uo501\#123 & 17:55:27.6343 $+$18:18:55.559 & 17.97 / 4.09 & 10.9 & 31.4 & B        & B  & \nd                      \\
       89 & uo501\nd   & 17:55:30.3430 $+$18:18:30.982 & 17.81 / 5.62 & 14.8 & 10.3 & B        & -  & \nd                      \\
       90 & uqc01\#065 & 18:07:04.9266 $+$45:44:13.350 & 18.57 / 4.09 &  8.3 & 20.4 & {\em BD} & DB & \nd                      \\
       91 & uqc01\#097 & 18:07:06.4575 $+$45:44:34.643 & 18.43 / 4.23 &  9.5 & 28.1 & BD       & B  & faint disk, 1bc (lensed?) \\
       92 & uqc00\#071 & 18:07:35.4057 $+$46:00:02.849 & 18.03 / 4.38 &  5.4 & 26.0 & ID       & D  & irr or asym disk + dust  \\
       93 & uqc00\#111 & 18:07:43.7618 $+$45:59:45.389 & 18.26 / 4.60 &  6.7 & 18.2 & B        & B  & faint disk?              \\
       94 & u2fq1\#130 & 21:53:30.6614 $+$17:41:45.643 & 17.91 / 4.05 &  5.1 & 48.9 & D        & D  & dust lane                \\
       95 & u2fq1\#389 & 21:53:32.4668 $+$17:41:28.683 & 18.39 / 5.25 &  5.6 &  5.6 & D        & D  & LSB                      \\
       96 & u2fq1\#166 & 21:53:32.4989 $+$17:42:53.920 & 18.02 / 4.98 &  6.9 & 15.2 & {\em DB} & BD & \nd                      \\
       97 & u2fq1\#273 & 21:53:33.0947 $+$17:42:52.077 & 18.15 / 4.43 &  5.9 & 21.7 & ID       & D  & 1bc, poss merger         \\
       98 & u2fq1\#158 & 21:53:33.3712 $+$17:42:49.533 & 17.96 / 4.45 &  7.4 & 19.2 & DB       & DB & asym disk, 1fc           \\
       99 & u2fq1\#128 & 21:53:33.8187 $+$17:41:15.458 & 17.37 / 4.10 & 11.7 & 63.4 & B        & DB & asym nucl, mbc, merger?  \\
      100 & u2fq1\#184 & 21:53:33.8420 $+$17:43:01.576 & 18.13 / 4.59 &  7.2 & 22.1 & DB       & DB & \nd                      \\
      101 & u2fq1\#095 & 21:53:34.0945 $+$17:42:40.723 & 16.77 / 5.02 & 26.6 & 53.8 & {\em DB} & BD & lg disk, 2bc             \\
      102 & u2fq1\#416 & 21:53:34.5140 $+$17:43:05.921 & 18.53 / 5.01 &  5.0 &  7.7 & U        & G  & asym, 1fc                \\
      103 & u2fq1\#107 & 21:53:38.5288 $+$17:42:18.217 & 17.17 / 4.24 & 14.6 & 59.8 & B        & B  & faint disk?              \\
      104 & u2fq1\#085 & 21:53:38.6836 $+$17:41:07.398 & 17.04 / 4.29 & 13.2 & 42.9 & {\em BD} & BD & bright S0/Sa?            \\
      105 & u2fq1\#115 & 21:53:38.8887 $+$17:42:25.657 & 17.86 / 4.31 & 11.6 & 30.5 & B        & B  & \nd                      \\
      106 & u2fq1\#099 & 21:53:39.1231 $+$17:42:25.960 & 17.51 / 4.17 & 16.2 & 59.8 & D        & DB & lg disk, asym            \\
      107 & u2fq1\#224 & 21:53:39.7393 $+$17:41:12.323 & 18.18 / 5.02 &  7.2 & 14.4 & {\em DB} & B  & \nd                      \\
      108 & u2v16\nd   & 22:17:35.4817 $+$00:17:34.004 & 18.82 / 4.60 &  5.9 & 10.8 & U        & -  & nearby comp.             \\
      109 & u2h91\#034 & 22:17:35.8459 $+$00:13:51.524 & 17.70 / 4.27 &  8.8 & 67.4 & DB       & DB & lg disk, poss bar, 2fc\tnm{f} \\
      110 & u2h91\#011 & 22:17:37.6791 $+$00:15:57.024 & 17.25 / 4.14 &  8.8 & 98.2 & {\em D}  & DB & lg asym disk; \tnm{f}    \\
      111 & u2fq2\nd   & 22:47:09.9607 $-$02:05:57.959 & 17.58 / 4.34 & 10.9 & 38.3 & D        & -  & lg disk w/knots, \tnm{g} \\
      112 & uec00\#057 & 23:04:24.4209 $+$03:04:10.043 & 18.19 / 4.15 &  5.7 & 23.1 & DB       & DB & \nd                      \\
      113 & uec00\#053 & 23:04:29.4158 $+$03:03:31.358 & 18.01 / 4.10 &  6.4 & 23.6 & DB       & BD & asym disk                \\
      114 & uec00\#080 & 23:04:30.4987 $+$03:04:43.728 & 18.07 / 4.20 &  6.7 & 14.6 & D        & D  & LSB disk or irr          \\
      115 & uec00\#047 & 23:04:31.2224 $+$03:04:35.336 & 17.81 / 4.38 &  7.4 & 16.5 & {\em D}  & BD & asym, dust               \\
   \enddata                                                                                  
   \tablenotetext{a}{Coordinates are in the format 00$^h$00$^m$00\fs00
      $\pm$00$^\circ$00$^\prime$00\farcs00, derived from the HST world
      coordinate system.}

   \tablenotetext{b}{Morphological classifications from this paper (YT) and
      the Medium Deep Surcey (MDS).  The primary classifications are (B)
      Bulge, (BD) Bulge-dominated disk, (DB) Disk with a bulge, and (D) Disk.
      See text for further details.}

   \tablenotetext{c}{Abbreviations in the comments: lg$=$large; 
      asym$=$asymmetric; 1fc, 2fc, mfc$=$[1,2,multiple] faint companions; 
      1bc, 2bc, mbc$=$[1,2,multiple] bright companions; LSB$=$Low Surface 
      Brightness; AGN$=$Active Galactic Nucleus; dbl$=$double; nucl$=$nucleus; 
      v$=$very; irr$=$irregular; poss$=$possible; comp$=$companion.}

   \tablenotetext{d}{Italicized entries indicate that the galaxy appears
      sufficiently inclined that dust may make a significant contribution 
      to the integrated color.}

   \tablenotetext{e}{The MDS incorrectly fits these galaxies with multiple
      components}

   \tablenotetext{f}{u2h91\#034 is SSA22 Hawaii\#77 at z=1.02; u2h91\#011
      is SSA22 Hawaii\#64 at z=0.653  }

   \tablenotetext{g}{This object could drop off the sample as some light
      from the extended disk is lost in the gap between WFPC2 CCDs.}

\end{deluxetable}
\clearpage


\onecolumn

\end{document}